\documentclass[sn-apa]{sn-jnl}

\usepackage{graphicx}%
\usepackage{multirow}
\usepackage{amsmath,amssymb,amsfonts}%
\usepackage{amsthm}%
\usepackage{mathrsfs}%
\usepackage[title]{appendix}%
\usepackage{xcolor}%
\usepackage{textcomp}%
\usepackage{manyfoot}%
\usepackage{booktabs}%
\usepackage{algorithm}%
\usepackage{algorithmicx}%
\usepackage{algpseudocode}%
\usepackage{listings}%
\usepackage{caption}%
\usepackage{subcaption}%

\theoremstyle{thmstyleone}%

\theoremstyle{thmstyletwo}%

\theoremstyle{thmstylethree}%

\begin{document}

\title[CycleGAN for MRI]{CycleGAN Models for MRI Image Translation}

\author*[1]{\fnm{Cassandra} \sur{Czobit}}\email{cczobit@torontomu.ca}

\author[1,2]{\fnm{Reza} \sur{Samavi}}\email{samavi@torontomu.ca}

\affil*[1]{\orgdiv{Electrical, Computer and Biomedical  Engineering}, \orgname{Toronto Metropolitan University}, \city{Toronto}, \state{Ontario}, \country{Canada}}

\affil[2]{\orgname{Vector Institute for Artificial Intelligence}, \city{Toronto}, \state{Ontario}, \country{Canada}}

\abstract{Image-to-image translation has gained popularity in the medical field to transform images from one domain to another. 
Medical image synthesis via domain transformation is advantageous in its ability to augment an image dataset where images for a given class is limited. From the learning perspective, this process contributes to data-oriented robustness of the model by inherently broadening the model's exposure to more diverse visual data and enabling it to learn more generalized features. In the case of generating additional neuroimages, it is advantageous to obtain unidentifiable medical data and augment smaller annotated datasets. This study proposes the development of a CycleGAN model for translating neuroimages from one field strength to another (e.g., 3 Tesla to 1.5).  This model was compared to a model based on DCGAN architecture. CycleGAN was able to generate the synthetic and reconstructed images with reasonable accuracy. The mapping function from the source (3 Tesla) to target domain (1.5 Tesla) performed optimally with an average PSNR value of 25.69 $\pm$ 2.49 dB and an MAE value of 2106.27 $\pm$ 1218.37. }





\maketitle

\section{Introduction}\label{sec1}
The field of image-to-image translation is increasingly prominent in enhancing the data-oriented robustness of machine learning methods, particularly in the realm of medical imaging where it facilitates the translation of images across different domains. This technique proves invaluable in medical image synthesis through domain conversion, especially useful when there is a scarcity of images for certain classes, which also aids in safeguarding patient privacy. From the learning perspective, such translation processes improves model robustness by data augmentation that results in a model's exposure to a wider array of visual data. Thus, the approach helps in fostering the development of more generalized feature recognition capability. From a medical imaging perspective, the technique is specifically instrumental in mild traumatic brain injuries (mTBIs), where it routinely becomes  difficult to diagnose due to a lack in quantitative assessments \cite{Aoki2012DiffusionMeta-Analysis}. 

Magnetic resonance imaging (MRI) is known universally as a form of technology to obtain images for disease diagnosis and prognosis, in a non-invasive manner \cite{You2022FineDomain}. Diffusion tensor imaging (DTI) has become a promising diagnostic tool as an MRI-based technology to evaluate the organization of white matter in neuroimaging \cite{Aoki2012DiffusionMeta-Analysis}. These tools, with machine learning, can be used to characterize, monitor and generate predictions on disease progression, while increasing efficacy and efficiency in patient care \cite{Hosny2018ArtificialRadiology}.

The optimal field strength for assessing DTI images has been frequently debated. The magnetic field strength of DTI neuroimages range from 0.2-0.5 Tesla (T) for low field images to 7T and above for high field imaging \cite{You2022FineDomain}. A field strength of 1.5T or 3T are predominantly selected for clinical MRI procedures, due to the high signal-to-noise ratio (SNR) and resolution \cite{Campbell-Washburn2019OpportunitiesMRI},\cite{Marques2019Low-fieldPerspective}. Higher field strengths have traditionally been viewed as advantageous due to an increase in image contrast and resolution \cite{Bahrami2016ReconstructionMRI},\cite{Hori2021Low-FieldRenaissance}. However, DTI at a field strength of 0.5T is considered advantageous for assessing head trauma, due to the reduction of geometric distortions and susceptibility artifacts \cite{Campbell-Washburn2019OpportunitiesMRI},\cite{Wiens2020Feasibility0.5T}.

Generative Adversarial Network (GAN) models have gained an increasing amount of popularity for medical image generation, reconstruction, and classification \cite{Yi2019GenerativeReview}, \cite{Kazeminia2020GANsAnalysis}. GAN models, such as Deep Convolutional GAN, are primarily used as a method for augmenting datasets. Composed of a generator and discriminator, the generator creates synthetic images from the probability distribution, whereas the discriminator classifies samples as real or fake \cite{Goodfellow2014GenerativeNetworks}. The generator obtains an input dimension $z$ from the Gaussian distribution to create the synthetic samples, $G(z)$, and maps the pseudo-sample distribution \cite{Lan2020GenerativeInformatics}. The ideal discriminator model will maximize the probability of allocating the correct label to the sample \cite{Goodfellow2014GenerativeNetworks}. Translation GAN models, such as CycleGAN, function similarly by using two GAN networks to generate target images from a source image. The creation of realistic synthetic medical images offers a solution to alleviate privacy concerns relating to diagnostic imaging and usage of individuals’ medical data \cite{Yi2019GenerativeReview}. 

The primary purpose of this investigation is to demonstrate a proof-of-concept for translating neuroimages between two field strengths in the MRI image domain. We assess two models for data transformation using two field strength and data augmentation for images in the same field strength. We develop a CycleGAN model for data transformation from a source domain to the target domain (i.e., 3T to 1.5T). Our second model aims to augment the 3T and 1.5T datasets by generating synthetic images from the same field. Our empirical evaluation shows that the CycleGAN model is able to translate images within the same modality, whereas the DCGAN model produces synthetic images of poor quality.

\section{Related Work}\label{sec2}
GANs were proposed as a mechanism to generate images that approximate the data distribution of the input functions \cite{Goodfellow2014GenerativeNetworks}. 
Thus, GANs have emerged as a powerful tool for generating synthetic data, which can be particularly useful for data augmentation in DNNs training. The usage of GANs for this purpose has been explored in various research studies and applications. For example,~\cite{Boursalie2021EvaluationModels} investigated GAN-based imputation models to address data imbalance and metrics to evaluate such models in contrast to the statistical imputation models.  

In image-to-image translation, adaptations of the original GAN model have addressed issues surrounding convergence of the models and mode collapse \cite{Yi2019GenerativeReview}. A deep convolutional GAN model (DCGAN) was created to overcome issues regarding the training stability by implementing fully convolutional upsampling/downsampling layers \cite{Yi2019GenerativeReview}, \cite{Radford2015UnsupervisedNetworks}.

Paired image translation requires data from both modalities \cite{Yi2019GenerativeReview}. A study by Emami et al., explored paired image translation from T1-weighted input MRI images to generate synthetic CT scans \cite{Emami2018GeneratingNetworks}. The conditional GAN model involved a residual network as the generator and a CNN for the discriminator. A model by Bahrami et al., constructed a Canonical Correlation Analysis space from paired 7T and 3T MRI images, where the extracted patches were mapped to a common space to show greater structural detail in the segmentation of white and gray brain matter \cite{Bahrami2016ReconstructionMRI}. Unpaired image translation offers more flexibility from paired images in instances where paired data or an abundance of labelled data is not available. A CycleGAN model was developed to utilize unpaired images, without a task specific similarity function, for image translation \cite{Zhu2017UnpairedNetworks}. The proposed mapping approach coupled the adversarial loss with inverse mapping, along with a cycle consistency loss function, to translate the images from domain X to Y, and vice versa \cite{Zhu2017UnpairedNetworks}. Hiasa et al., implemented a CycleGAN model for cross-modality image translation between CT and MRI scans and adapted the loss function to capture variation of intensity pairs in the images \cite{Hiasa2018Cross-modalitySize}.  

The proposed study differs from prior work, in which the input and output images will be within the same MRI domain, but at varying field strengths. Previous investigations have translated images across modalities (e.g., MRI to CT images), but have not addressed the translation of field strengths within a single modality. This translation aims to visualize the MRI image from a different anatomical granularity level. 

\section{CycleGAN Image Translation Model}\label{sec4}
The models discussed in the following sections were developed for the purpose of image-to-image translation from 3T to 1.5T to generate synthetic neuroimages. It consists of 2 mapping functions, G and F, with a relationship defined as $G^*,F^*=\text{arg min max }\mathcal{L}(G,F,D_X,D_Y)$ \cite{Zhu2017UnpairedNetworks}. Section 3.1 details how the data was obtained, section 3.2 describes the architecture of the CycleGAN model, and section 3.3 describes the DCGAN network for data augmentation. 

\subsection{Data}
Two field strengths, 1.5T and 3T, were selected as part of this study for DTI image translation. For the purpose of this study, these field strengths were selected to demonstrate a proof of concept for the CycleGAN and DCGAN models. DTI images at a field strength of 0.5T are not widely available for public use. For both datasets, 70\% of the images were assigned for training purposes, with the remaining 30\% for testing. The scale of both datasets are small due to limited open-source resources for the required field strengths.

The 3T images were collected by the University of North Carolina-CH and were made available on the Kitware Data open-source platform.\footnote{\url{https://data.kitware.com/\#collection/591086ee8d777f16d01e0724/folder/58a372fa8d777f0721a64dfa}} The scans were acquired on a 3T Siemens MR unit. The DTI images were obtained using 6 directions with 2mm x 2mm x 2mm isotropic resolution (voxel size). The 35 axial cross-sectional images were obtained from individuals in age groups from 18-29, 30-39, 40-49, 50-59, and 60+ years. From the 35 patient scans, ten slice were obtained from each brain volume, for a total of 350 slices.

The 1.5T scans were obtained on the Philips Intera MRI scanner from the SCA2 Diffusion Tensor Imaging dataset on the OpenNeuro platform.\footnote{\url{https://openneuro.org/datasets/ds001378/versions/00003}} Diffusion weighted images were collected for 16 individuals for 2 sessions, a year apart. The axial images were applied in 15 directions for 50 slices, with a slice thickness of 3 mm. For each image, ten slices were obtained to represent the 2-dimensional DTI scan, for a total of 16 images and 160 slices.

\subsection{CycleGAN}
The proposed architecture for the CycleGAN model consists of two sequential models that operate in a forward and backward cycle for images in the source and target domains. Each model contains a generator and discriminator component. The complete CycleGAN network translates an image in source domain to the target domain, where the source represents the 3T images, and target contains the 0.5T images.

Within this network, the mapping functions between the two domains are trained for the two data distributions. Two mapping functions are defined as G: X → Y and F: Y → X, where X is the training sample in the source domain and Y is the training sample in the target domain \cite{Zhu2017UnpairedNetworks}. The adversarial discriminator for each mapping function contains terms for the adversarial and cycle consistency loss functions.

In the discriminator model, the CNN model architecture classifies the images and is identical for each of the mapping functions, G and F. The discriminator receives images of size 256 x 256 from the generator and source domain as the input. We use a 70 x 70 PatchGAN method for patch predictions. The PatchGAN evaluates NxN pixels to classify as real or fake, and averages the score across the entire image \cite{Zhu2017UnpairedNetworks}. This approach reduces the number of parameters from a full-image discriminator \cite{Zhu2017UnpairedNetworks}. The discriminator consists of 5 convolutional layers, stride 2 and $k$ filter size, where $k$ is equal to the number of filters per layer. The value of $k$ per layer is 64, 128, 256, 512 and 1, for layers 1 to 5 respectively. We apply a 4x4 filter for filters 1-4, followed by a LeakyReLU activation function with a slope of 0.2. Instance normalization is applied to layers 2, 3 and 4.

The generator architecture is constructed as an encoder-decoder network with a residual network, ResNet, to transform the input images. The encoder portion of the network downsamples the image, and consists of 3 convolutional layers, with a ReLU activation function, followed by instance normalization. The ResNet receives the output of the encoder as the input. ResNet contains 9 convolutional blocks, which functions by creating skip connections between layers, where a residual block has a layer feeding into another 2-3 skips away in addition to the layer succeeding it \cite{He2015DeepRecognition}. From the ResNet network, the output is passed to the decoder, where upsampling is performed using two deconvolution blocks with fractional stride values to generate the final output as the original size.

The complete composite model of the generator and discriminator is compiled for both mapping functions, and trained for 50 epochs, with a batch size of 4. The Adam optimizer is selected with a learning rate of 0.0002 and a momentum value of 0.5 \cite{Zhu2017UnpairedNetworks}. We trained the complete model from scratch.

\subsection{DCGAN}
The DCGAN model consists of a generator and discriminator model for each field strength. This model generates synthetic images from a singular source domain for the respective field strength (i.e., a separate model for 3T and 1.5T). For both DCGAN models, the generator and discriminator architectures are constructed in the same manner. The DCGAN architecture was adapted from Radford et al. \cite{Radford2015UnsupervisedNetworks}.

The discriminator architecture contains 3 convolutional layers with a 4x4 kernel and stride 2 in each layer for downsampling. For each layer, the LeakyReLU activation function is used, followed by a dropout rate of 0.4. The convolutional layers are flattened, before being passed to a Dense layer, with the sigmoid activation function.

The generator is composed of an initial Dense layer, followed by 3 transpose convolutional layers. The Dense layer receives a low-resolution version of the image, which is followed by the
LeakyReLU activation function with alpha equal to 0.2. The transpose convolution layers have filters of 256, 128 and 64 respectively with a 4x4 kernel size. The LeakyReLU activation function with an alpha value of 0.2 is applied to each layer. The final layer is a convolutional layer with a 3x3 kernel and uses the sigmoid activation function. 

The complete model is compiled and trained for 50 epochs, with a batch size of 4 and uses the binary cross-entropy loss function. The model is trained with the Adam optimizer, using a learning rate of 0.0002 and momentum of 0.05. 

\subsection{Evaluation}
Several statistical measures will be reviewed to evaluate the performance of the GAN variant models. Evaluation measures include mean absolute error (MAE), mean squared error (MSE) and peak signal-to-noise ratio (PSNR) \cite{Wolterink2017DeepData},\cite{Li2021WhenOpportunities}. To evaluate the image quality, we compared the synthesized 3T and 1.5T neuroimages to the real, ground truth images.

\section{Results and Discussion}\label{sec5}
\subsection{CycleGAN}
The CycleGAN model can be evaluated based on a visual inspection of the translation process, in addition to its reconstructive capabilities. One randomly selected ground truth image from the source domain was transformed to generate an image in the target domain, before being reconstructed back to the source image. Figure 1 (a) displays the translation process for the 3T image to 1.5T at 50 epochs, while Figure 1 (b) shows the image translation from 1.5T to 3T.

\begin{figure}
     \centering
     \begin{subfigure}[b]{0.75\textwidth}
         \centering
         \includegraphics[width=\textwidth]{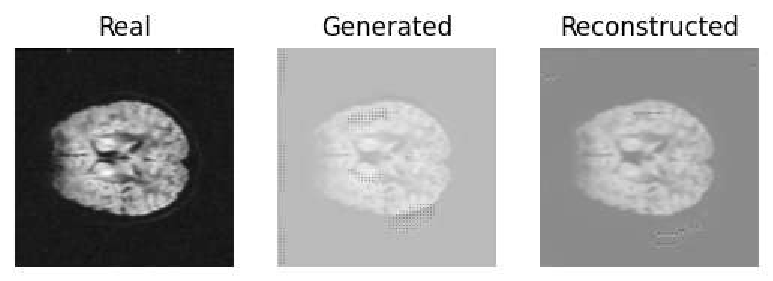}
         \caption{}
         \label{fig:y equals x}
     \end{subfigure}
     \hfill
     \begin{subfigure}[b]{0.75\textwidth}
         \centering
         \includegraphics[width=\textwidth]{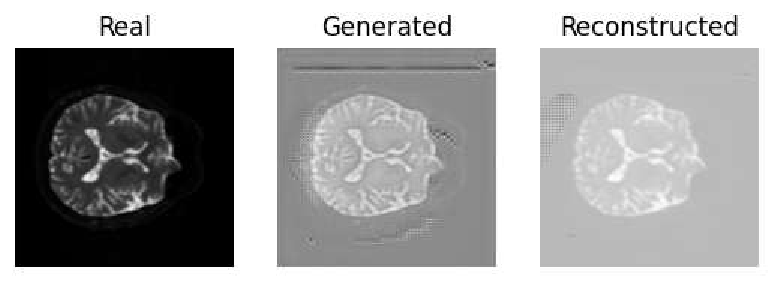}
         \caption{}
         \label{fig:three sin x}
     \end{subfigure}
        \caption{CycleGAN Generated and Reconstructed Images at 50 Epochs from 3T to 1.5T to 3T (a) and from 1.5T to 3T to 1.5T (b)}
        \label{fig:cycleGAN translation}
\end{figure}

From a visual perspective, it appears that both GAN models performed similarly. We performed a quantitative comparison of the image quality metrics for both models at 50 epochs. One thousand images were randomly selected to compute the reconstructive quality between the original, real image and the reconstructed image. These metrics were averaged for the translation of the source domain to the target domain, shown in Table 1 of Appendix A. The average value of the MAE, MSE and PNSR metrics were 2106.27 $\pm$ 1218.37, 0.0.00328 $\pm$ 0.0032, and 25.69 db $\pm$ 2.49, respectively. 

In contrast, the backward cycle model translated the target domain images to the source domain. The average calculated value for MAE was 602.27 $\pm$ 147.41 and the MSE value was 0.00189 $\pm$ 0.0013. The PSNR had an average value of 27.22 db $\pm$ 0.30, as shown in Table 1 of Appendix A. 

It is evident that the backward cycle consistency function performed better than the forward cycle model, as indicated by the higher PSNR value. In addition, the backward cycle consistency loss function scored a lower value for the MAE and MSE measures, which indicates a higher model accuracy. 

\subsection{DCGAN}
The DCGAN model was evaluated for the 1.5T image synthesis. The synthesized images are shown in Figure A1 in Appendix A and are evaluated based on a visual inspection and image quality metrics. Visually, it appears that the model was unable to generate a diverse set of examples. From 1000 randomly generated synthetic images, the average PSNR and MAE values were 6.20 db $\pm$ 0.11 and 31907.44 $\pm$ 415.85, respectively. The MSE was calculated as 0.24 $\pm$ 0.0059. This quantitative assessment demonstrates that the model was not able to generate high-quality synthetic images. This type of instability is an indicator of mode collapse, wherein the generator cannot create a diverse set of images \cite{Goodfellow2016NIPSNetworks}, \cite{Kodali2017OnGANs}. As a result, the discriminator only learns from a small subset of images \cite{Kodali2017OnGANs}. Increasing the size of the training set could be helpful to mitigate low accuracy results \cite{Lan2020GenerativeInformatics}.

We then evaluated the quality of synthesized 3T scans generated by the DCGAN model. This was performed for comparison purposes with the backward cycle consistency loss function of the CycleGAN model. From a qualitative perspective, the images are able to replicate some details, however, they lack clarity overall. These images are shown in Figure A2 of Appendix A. Quantitatively, evaluation metrics were calculated from 1000 synthesized images. The 3T DCGAN model generated images with a higher PSNR value than the 1.5T counterpart, with an average of 8.86 db $\pm$ 0.76. 

\subsection{Comparison of Models}
A comparison of the quantitative evaluation measures demonstrates the feasibility for using CycleGAN to augment datasets and translate MRI images. In contrast, the DCGAN model was not able to generate a diverse set of realistic synthetic images. These findings are evidenced by the higher PSNR values and lower error values (MAE and MSE) for the CycleGAN models.

\subsection{Research Limitations}
The findings of this investigation are limited by the small dataset for each domain. As opposed to selecting a range of 10 image slice of the DTI scans, a future study could input several different sizes of slices from each patient scan to increase the variability of training images for the CycleGAN and DCGAN models. 

\section{Conclusions}
The purpose of this investigation was to determine if the CycleGAN model could be applied within a single modality to translate MRI images between two field strengths. This investigation demonstrated that the application of this model has potential, barring some modifications to the dataset. In contrast, the DCGAN model failed to produce realistic synthetic images for both domains.

Future investigations could implement an additional model to review the quality of the images. The creation of a classification model could play a role in determining if the synthesized images are able to be correctly assigned to the desired category \cite{Yi2019GenerativeReview}. This addition would provide more concrete evidence for the performance of the CycleGAN model. Increasing the size of the experiment to include additional datasets and a larger scale of data would strengthen the current findings as well.

\bmhead{Acknowledgments}
We would like to thank Dr. Michael Noseworthy and Nicholas Simard for the initial discussions on current trends and limitations in medical imaging and motivating this problem.

\bibliography{references}

\begin{appendices}
\section{}\label{secA1}
\begin{table}[ht]
\centering
\caption{Comparison of Image Quality Metrics Across 1000 Synthesized Images}
\begin{tabular}[t]{lcccc}
\hline
Model &MAE $\pm$ SD &MSE $\pm$ SD &PSNR (dB) $\pm$ SD\\
\hline
CycleGAN 3T to 1.5T& 2106.27 $\pm$ 1218.37& 0.00328 $\pm$ 0.0032& 25.69 $\pm$ 2.49 \\
CycleGAN 1.5T to 3T& 602.27 $\pm$ 147.41& 0.00189 $\pm$ 0.0013& 27.22 $\pm$ 0.30\\
DCGAN 1.5T& 31907.44 $\pm$ 415.85&0.24 $\pm$ 0.0059 & 6.20 $\pm$ 0.11 \\
DCGAN 3T& 22559.57 $\pm$ 1875.35&0.14 $\pm$ 0.021 & 8.68 $\pm$ 0.76 \\
\hline
\end{tabular}
\end{table}%

\begin{figure}[ht]
\begin{center}
\includegraphics[width=0.65\textwidth]{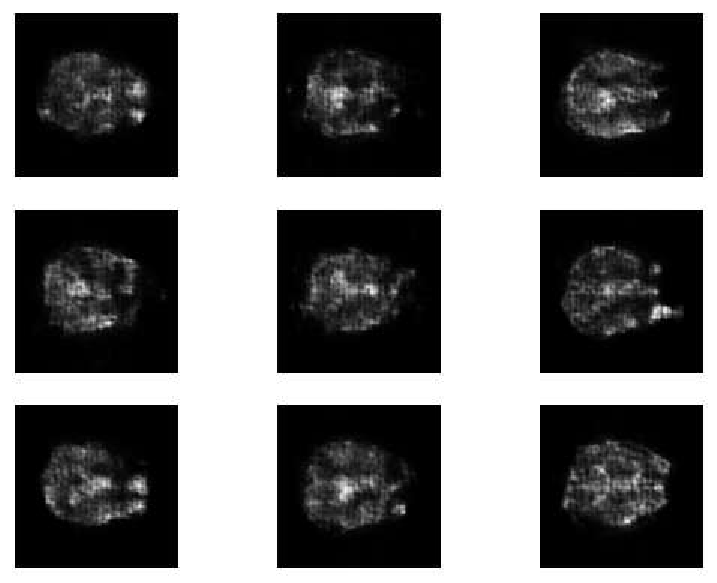}
\caption{Generation of Synthetic 1.5T DTI Scans After 50 Epoch}
\end{center}
\end{figure}

\begin{figure}[t]
\begin{center}
\includegraphics[width=0.65\textwidth]{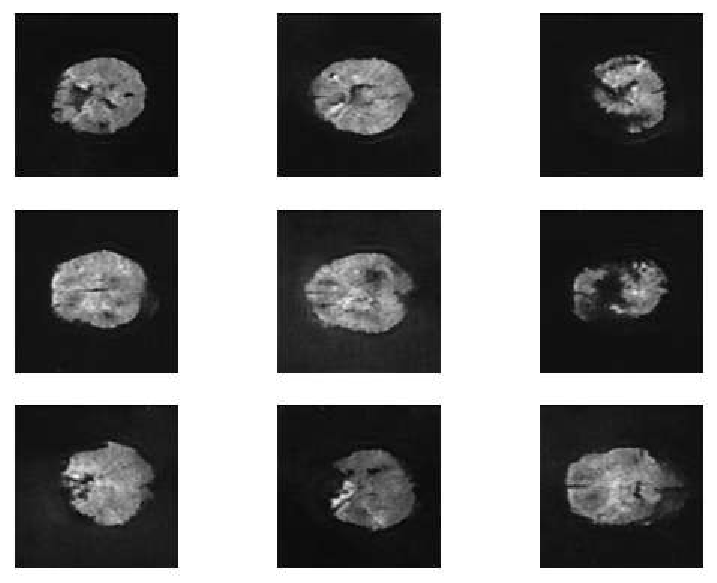}
\caption{Generation of Synthetic 3T DTI Scans After 50 Epoch}
\end{center}
\end{figure}

\end{appendices}

\end{document}